\documentstyle[epsf]{elsart}

\begin{document}
\begin{frontmatter}

\title{ SU(3) Symmetry and Scissors Mode Vibrations in
Nuclei }
\author{ Yang Sun$^{(1,2,3)}$, Cheng-Li Wu$^{(4)}$,
Kumar Bhatt$^{(5)}$, Mike Guidry$^{(1)}$ }
\address{
$^{(1)}$Department of Physics and Astronomy, University
of Tennessee\\ Knoxville, Tennessee 37996, USA \\
$^{(2)}$Department of Physics, Tsinghua University,
Beijing 100084, P. R. China \\
$^{(3)}$Department of Physics, Xuzhou Normal University\\ Xuzhou,
Jiangsu 221009, P. R. China \\
$^{(4)}$National Center for Theoretical Science,
Hsinchu, Taiwan 300, ROC \\
$^{(5)}$Department of Physics and Astronomy, University
of Mississippi\\ University, Mississippi 38677, USA \\
}
\maketitle

\begin{abstract}
We show that a nearly perfect $SU(3)$ symmetry emerges from an
extended Projected Shell Model. Starting from a deformed
potential we construct separate bases for neutron and proton
collective rotational states by exact angular momentum
projection. These rotational states are then coupled by
diagonalizing a residual pairing plus quadrupole interaction. The
states obtained exhibit a one-to-one correspondence with an
$SU(3)$ spectrum up to high angular momentum and excitation, and
their wave functions have a near-maximal overlap with the $SU(3)$
states. They can also be classified as rotational bands built on
spin-1$\hbar$ phonon excitations, which correspond to a
geometrical scissors mode and its generalizations. This work is a
direct demonstration that numerical angular momentum projection
theory extends the Elliott's original idea to heavy nuclear
systems.
\end{abstract}
\end{frontmatter}

\section{Introduction}

In the long history of the shell model, Elliott was the
first to point out the advantage of using a deformed
(intrinsic) many-body basis and developed the $SU(3)$
Shell Model \cite{Elliott} for $sd$-shell nuclei. In
this model, the classification of basis states and their
projection onto good angular momentum can be carried out
using group theory. It works nicely as long as the
spin-orbit force is weak. In heavier nuclei, where the
presence of a strong spin-orbit force is essential for
the correct shell closures, the high-$j$ orbital in an
$N$-shell is strongly pushed down by this interaction
and intrudes into a lower $N$-shell. This results in a
re-classification of basis states and the Elliott
$SU(3)$ scheme is no longer valid for heavy nuclei.

However, in early work of Bhatt, Parikh and McGrory
\cite{Bhatt}, it was shown that wave functions for
yrast states of deformed nuclei in the $1f$-$2p$ shell
that were
obtained in shell model
calculations within a severely truncated configuration
space that cannot support microscopic $SU(3)$ symmetry
of Elliott, have $SU(3)$-like structures when the wave
function of a state with total angular momentum $I$ is
expressed in terms of coupling of the collective neutron
and proton angular momenta $I_\nu$ and $I_\pi$ within the
truncated configuration space. On the basis of this
analysis, they suggested that the rotational features of
yrast states in heavy, deformed nuclei could be due to a
similar manifestation of $SU(3)$ symmetry occurring at
the macroscopic level, even if it is absent at the single
particle level in the usual basis.
Brink et al. \cite{Br87}  suggested an alternative derivation
without the use of symmetry arguments as well as the assumption of the
motion of a particle in a static deformed field.  Instead they approached the
problem by diagonalising a rotationally invariant Hamiltonian using basis
states of good angular momentum at all stages, which leads to the standard
`symmetrised' wavefunctions.

In our recent letter \cite{rotor.97}, the connection
between the numerically calculated states of heavy
nuclear system and such macroscopic $SU(3)$ symmetry
states has been explored. The work was based on an
extended Projected Shell Model (PSM). Starting from a
deformed potential, instead of employing only one BCS vacuum as in
the original Projected Shell Model \cite{HS.95}, we
use independent BCS condensates for neutrons and protons
to construct separate bases for neutron and
proton collective rotational states by exact angular
momentum projection. The active nucleons in the model space
were permitted to occupy three full major shells, which is a
space far larger than a conventional shell model can
handle. These rotational states were then coupled by
diagonalizing a residual pairing plus quadrupole
interaction. Many new bands emerge that are not
contained in the original PSM, and it was shown that these
bands exhibit a nearly perfect $SU(3)$ spectrum, even
though there is no explicit dynamical symmetry in such a
model.

Our work \cite{rotor.97} demonstrates a clear
connection between a numerical microscopic method, which starts
from certain set of intrinsic states, and an elegant
group theoretical description of nuclear states.  Thus the
PSM may be viewed as a natural method to
investigate the emergence of
$SU(3)$ symmetry as a truly dynamical symmetry in
heavy, deformed nuclear systems.

The present paper elaborates on our previous letter
\cite{rotor.97}. The paper is organized as follows: In Section
2, we introduce the extended PSM with an emphasis on how to
treat coupled rotations of neutron and proton systems. It is
shown that this treatment generalizes the original PSM by
allowing for relative motion between the neutron and proton
systems. In Section 3, we give details of the calculation for an
example, $^{168}$Er. A much richer spectrum is obtained after
diagonalization than from the usual PSM. With the help of group
theory, the states obtained from the diagonalization are shown to
possess a strong $SU(3)$ symmetry. We further demonstrate that
these states can be classified as rotational bands built on
spin-$1\hbar$ phonon excitations, and we suggest that these
correspond to a geometrical scissors mode and its
generalizations. These states are interpreted in Section 4. In
Section 5, we buttress our conclusions by comparing the angular
momentum distribution in our deformed basis states and the
intrinsic states of the  $SU(3)$ model, and by evaluating
overlaps of our total wave functions with the $SU(3)$ ones. Then
we calculate $B(E2)$ values from the PSM wave functions to
further confirm the state classifications. Some comments about the
emergent $SU(3)$ symmetry and about a possible relation to
scissors mode vibrations are made in Section 6 and 7,
respectively. The lowest excitation bands, the first 1$^+$ and the
first 2$^+$ band, as well as the second 0$^+$ band, are further
discussed in Section 8. Finally, we summarize the paper in
Section 9.

\section{Extension of the Projected Shell Model}

The shell model is the most fundamental way of describing
many-nucleon systems fully quantum mechanically. However, using
the shell model to study deformed heavy nuclei is a desirable but
very difficult task because of large dimensionality and its
related problems. Even with today's computer power, the best
standard shell model diagonalization can be done only in the full
$pf$-shell space, as for example in the work of the
Strasbourg--Madrid group \cite{Zuker}, for which the dimension of
the configuration space is well over one million. The PSM
provides one possible solution for this difficulty. In this
approach, one first truncates the configuration space with
guidance from the deformed mean field by selecting only the BCS
vacuum plus a few quasiparticle configurations in the Nilsson
orbitals around the Fermi surface, performs angular momentum
projection (and particle number projection if required) to obtain
a set of laboratory-frame basis states, and finally diagonalizes
a shell-model Hamiltonian in this space. Since the deformed mean
field + BCS vacuum already incorporates strong particle-hole and
particle-particle correlations, this truncation should be
appropriate for the low-lying states dominated by quadrupole and
pairing collectivity. Indeed, this approach has been very
successful for ground band (g-band) properties and near-yrast
quasiparticle excitations in high-spin physics for both normally
deformed and superdeformed states \cite{HS.95,SD1,SD2,SD3}.

However, in this formalism the basis vacuum is the usual BCS
condensate of neutrons and protons in a fixed deformed
potential (fixed shape in the geometric models). Without
quasiparticle excitations, one obtains only the g-band
after angular momentum projection. Thus, there is no
room for studying any other collective excitations.

It is known that small perturbations of nuclear shapes
and relative orientations around the equilibrium can
give rise to physical states at low to moderate
excitation energies. Classical examples of such motion
include $\beta$- and $\gamma$-vibrations \cite{BMbook},
in which neutrons and protons undergo vibrations as a
collective system. These small-amplitude motions are not
built into the ground state for theories like
Hartree--Fock--Bogoliubov or BCS. One may obtain the
$\beta$- and $\gamma$-vibrations
by building additional correlations into the ground state, as
for example in the Random Phase Approximation (see, for
example, Ref. \cite{RPA}),
or by enriching the intrinsic degrees of freedom in the deformed
potential \cite{gamma}.

Another possible mode of collective motion results if we consider
neutron and proton fields of the fixed deformation, but permit
small oscillations in the relative orientations of these fields
\cite{Rowe,Hilton}. This geometric picture may be related to the
two-rotor model \cite{two.rotor}. Because of the strong restoring
force \cite{nojarov.86} (with its physical origin in the
neutron-proton interaction), this oscillation is confined to a
small angle between the protons and neutrons (thus it is termed
small-amplitude scissors motion). This vibration, and the
$\beta$- and $\gamma$-vibrations, can be classified using group
theoretical methods;  they belong to the lowest collective
excitations of the ground state, as pointed out by Iachello
\cite{IBM}.

As introduced in our previous letter \cite{rotor.97}, in
order to study microscopically the relative motion
between neutron and proton intrinsic states with a fixed
deformation, we have extended the original PSM in the
following manner: Instead of a single BCS vacuum, the
angular momentum projection is now performed for
separate neutron and proton deformed BCS vacua. Although
the introduction of two separately projected BCS vacua
seems to treat neutrons and protons as two independent
systems, the equal deformation used in the Nilsson
calculation of the basis already contains strong
correlation between the two systems \cite{np-force}.
Further proton-neutron correlations are introduced by
explicitly diagonalizing the total Hamiltonian in the
basis formed by the angular momentum projected neutron
and proton states. This procedure gives the
usual g-band corresponding to the coherently coupled BCS
condensate of neutrons and protons, but also leads to a
new set of states built on a more
complex vacuum that incorporates fluctuations in the
relative orientation of the neutron and proton fields of constant deformation.

The neutron and proton valence spaces employed in the
present work are those of Ref.\ \cite{HS.95}. Our single
particle space contains three major shells ($N$ = 4, 5,
and 6) for neutrons and ($N$ = 3, 4 and 5) for protons;
this has been shown to be sufficient for a
quantitative description of rare-earth g-bands and bands
built on a few quasiparticle excitations \cite{HS.95}.
Separable forces of pairing plus quadrupole type
\cite{PQmodel} are used in our Hamiltonian.
We note that a recent study by Dufour and Zuker \cite{Zuker.96}
has shown explicitly that the residual part of the
realistic force is strongly dominated by pairing and
quadrupole interactions.

The Hamiltonian \cite{HS.95} can be expressed as $\hat
H = \hat H_\nu + \hat H_\pi + \hat H_{\nu\pi}$, where
$H_\tau$ $(\tau=\nu,\pi)$ is the like-particle  pairing plus quadrupole
Hamiltonian, with inclusion of quadrupole-pairing,
\begin{equation}
\hat {H}_{\tau}~=~\hat H^0_{\tau} -{\chi_{\tau\tau}\over
2} \sum_\mu \hat Q^{\dagger\mu}_{\tau} \hat
Q^{\mu}_{\tau} - G_{\mbox{\scriptsize M}}^{\tau}\hat
P^\dagger_{\tau} \hat P_{\tau} - G_{\mbox {\scriptsize
Q}}^{\tau} \sum_\mu \hat P^{\dagger\mu}_{\tau} \hat
P^{\mu}_{\tau},
\label{E}
\end{equation}
and $\hat H_{\nu\pi}$ is the n--p
quadrupole--quadrupole residual interaction
\begin{equation}
\hat H_{\nu\pi} ~=~ - \chi_{\nu\pi} \sum_\mu \hat
Q^{\dagger \mu}_\nu \hat Q^\mu_\pi.
\label{QQ}
\end{equation}
In Eq.\ (\ref{E}), the four terms are,
respectively, the spherical single-particle energy 
which contains a proper spin-orbit force \cite{NKM}, the
quadrupole-quadrupole interaction, the monopole-pairing,
and the quadrupole-pairing interaction. The interaction
strengths
$\chi_{\tau \tau}$ ($\tau = \nu$ or $\pi$) are related
self-consistently to the deformation $\varepsilon$ by
\cite{HS.95}
\begin{equation}
\chi_{\tau \tau} ~=~ {{{2\over 3}
\varepsilon (\hbar\omega_\tau )^2} \over
{\hbar\omega_\nu \langle \hat Q_0 \rangle_\nu +
\hbar\omega_\pi \langle \hat Q_0 \rangle_\pi}}.
\label{chi}
\end{equation}
Obviously, neutrons and
protons are coupled by the self-consistency condition.
Following Ref.\ \cite{HS.95}, the strength
$\chi_{\nu\pi}$ is assumed to be $\chi_{\nu\pi} ~=~
(\chi_{\nu\nu} \chi_{\pi\pi})^{1/2}$. Similar
parameterizations were used in earlier work
\cite{PQmodel}.

In this paper, we concentrate on
collective motion only and do not consider
quasiparticle excitations;  however, we shall discuss
possible physical implication of
including quasiparticle states in a later
section. Particle number projection is not included, so
particle number is conserved only on average in the
BCS calculation. It has been shown by a systematic
calculation \cite{HS.95} that particle number projection
does not alter the unprojected results very much for
those well-deformed heavy nuclei. We do not
expect that particle number projection will change
conclusions of this paper with respect to the emergence of
$SU(3)$ symmetry.

The intrinsic state $| 0 \rangle$ of an axially deformed nucleus is,
in this approximation, taken to be the product of the
Nilsson-BCS quasiparticle vacuum states $| 0_\nu \rangle$ and
$| 0_\pi \rangle$
\begin{equation}
| 0 \rangle ~=~ | 0_\nu \rangle | 0_\pi \rangle.
\label{product}
\end{equation}
The original PSM ground band with angular momentum $| I \rangle$
is obtained by angular momentum
projection \cite{HS.95} onto the vacuum:
\begin{equation}
| I \rangle ~=~
\aleph^I \hat P^I | 0 \rangle,
\label{ansatz}
\end{equation}
where $\hat P^I$ is the (one-dimensional) angular momentum
projection operator \cite{RSbook}
\begin{equation}
\hat P^I ~=~ (I + {1\over 2})
\int_0^\pi ~ d\beta~ sin\beta~ d^I(\beta)~ \hat R(\beta),
\end{equation}
with
$d^I(\beta)$ the small $d$-function and $\hat R(\beta)$ the
one-dimensional rotational operator \cite{Ebook} and $\aleph^I$ is the
normalization constant,
\begin{equation}
\aleph^I ~=~ \langle 0 | \hat P^I | 0 \rangle ^{-{1\over 2}}.
\end{equation}

In the extended PSM, we first project out the neutron (proton) states
$| I_\nu \rangle$ ($| I_\pi \rangle$) with angular momentum $I_\nu$ ($I_\pi$)
from the intrinsic state $| 0_\nu \rangle$ ($| 0_\pi \rangle$).
The projected states $| I_\nu \rangle$ and $| I_\pi \rangle$
are coupled to form the basis states $| ( I_\nu \otimes I_\pi) I \rangle$
for different total angular momentum $I$.
These basis states are used to construct the matrix of the total
Hamiltonian of Eqs.\ (\ref{E}) and (\ref{QQ}),
\begin{eqnarray}
\langle ( I_\nu \otimes
I_\pi) I | &\hat H& | ( I^\prime_\nu \otimes I^\prime_\pi) I \rangle
\nonumber\\
&~=~&\left [ \langle I_\nu | \hat H_\nu | I_\nu^\prime \rangle + \langle I_\pi
|
\hat H_\pi | I_\pi^\prime \rangle \right ]
\delta_{I_{\nu}I_\nu^\prime}\delta_{I_{\pi}I_\pi^\prime}
\nonumber\\
&-&
\chi_{\nu\pi} \langle ( I_\nu \otimes I_\pi) I
| \hat Q^\dagger_\nu \hat Q^{}_\pi | (
I^\prime_\nu \otimes I^\prime_\pi) I \rangle .
\label{energy}
\end{eqnarray}
The last term in Eq.\ (\ref{energy}) can be written explicitly as \cite{BMbook}
\begin{eqnarray}
\langle ( I_\nu &\otimes &I_\pi) I | \hat Q^\dagger_\nu \hat Q^{}_\pi
| ( I^\prime_\nu \otimes I^\prime_\pi) I \rangle
\nonumber\\
&~=~& {\cal W} ( I_\pi 2
I I^\prime_\nu; I^\prime_\pi I_\nu )~~ \langle I_\nu \parallel \hat Q_\nu
\parallel I^\prime_\nu \rangle
\nonumber\\
&\times&\langle I_\pi \parallel \hat
Q_\pi \parallel I^\prime_\pi \rangle / \sqrt{(2I_\nu+1)(2I^\prime_\pi+1)} ,
\end{eqnarray}
where $\cal W$ is the 6-$j$ symbol.

The term $\langle I_\nu | \hat H_\nu | I_\nu \rangle$
($\langle I_\pi | \hat H_\pi | I_\pi \rangle$) is the
energy of the state $| I_\nu \rangle$ ($| I_\pi
\rangle$) projected from the intrinsic state $| 0_\nu
\rangle$ ($| 0_\pi \rangle$) with the neutron (proton)
part of Hamiltonian $\hat H_\nu$ ($\hat H_\pi$) given by
Eq. (\ref{E}). The Hamiltonian matrix of Eq.
(\ref{energy}) is diagonalized and the resulting PSM
eigenstates $|\alpha, I \rangle_{\mbox{\scriptsize
PSM}}$ are expressed as a linear combination of the
basis states $| ( I_\nu \otimes I_\pi) I \rangle$:
\begin{equation}
|\alpha, I \rangle_{\mbox{\scriptsize
PSM}} ~=~ \sum_{I_\nu I_\pi} f^\alpha_{\mbox{\scriptsize
PSM}} (I_\nu I_\pi ; I) |( I_\nu \otimes I_\pi) I
\rangle ,
\end{equation}
where $\alpha$ labels different eigenstates
having the same angular momentum. We shall show in
the following sections that the amplitudes
$f^\alpha_{\mbox{\scriptsize PSM}} (I_\nu I_\pi ; I)$
carry information about $SU(3)$ symmetry emerging
dynamically from the extended PSM.

\section{An Example: The Spectrum of $^{168}$Er}

Let us take as a typical example the rotational nucleus,
$^{168}$Er. The deformed basis is constructed at deformation
$\varepsilon_2 = 0.273$ for this nucleus \cite{BFM.86}. The
numerical values of the parameters in the Hamiltonian (Eqs.
(\ref{E}) and (\ref{QQ})) appropriate for this nucleus are (in
MeV): $\chi_{\nu\nu} = 0.0206$, $\chi_{\pi\pi} = 0.0160$,
$\chi_{\nu\pi} = 0.0182$, $G^\nu_M = 0.1049$, and $G^\pi_M =
0.1346$. The ratio of the strength of the quadrupole-pairing
interaction to that of the monopole-pairing interaction is 0.16.
All these values are the same as those employed in the early
paper of Hara and Iwasaki \cite{HI.80} and as those in the
review article of Hara and Sun \cite{HS.95}. The states $|I_\nu
\rangle$ ($|I_\pi \rangle$) with angular momenta $I_\nu = 0, 2,
..., 32$ ($I_\pi = 0, 2, ..., 16$) were projected from the
Nilsson-BCS vacuum state $| 0_\nu \rangle$ ($| 0_\pi \rangle$).
We have omitted in this calculation the neutron states with
$I_\nu > 32$ and the proton states with $I_\pi > 16$ because
their probabilities in the intrinsic states are very small ($<$
0.001).

The energy spectra
$E(I_\tau) = \langle I_\tau | \hat H_\tau | I_\tau \rangle$
obtained from the projected states $|I_\tau \rangle$ are almost
rotational ( $E(I_\tau) = A_\tau I_\tau (I_\tau + 1)$ ),
with the moment of inertia parameters
$A_\nu =$ 0.019 MeV and $A_\pi =$ 0.048 MeV.

For each spin
$I$, we diagonalize Eq.\ (\ref{energy}) in the basis $|
( I_\nu \otimes I_\pi) I \rangle$. The
resulting spectrum with energy up to 20 MeV and angular
momenta up to 12$\hbar$ is shown in Fig.\ 1. It
shows strikingly regular structure.

The lowest band is the g-band, which is nearly identical
to that of earlier PSM calculations
\cite{HS.91} with the same Hamiltonian where no
separation and re-coupling of neutron and proton
components was considered. The calculated g-band in Fig.\
1 reproduces the experimental g-band \cite{BE2} very
well, as we shall show later in Fig.\ 4. In
addition to the g-band, many new excited bands emerge
that are not found in the earlier PSM calculations.
These bands exhibit a curvature similar to the g-band,
suggesting that their moments of inertia are similar to
the g-band.

\section{Emergence of SU(3) Symmetry}

The strikingly regular pattern shown in Fig.\ 1 can be
understood as manifestation of a nearly perfect
$SU(3)$ symmetry: all bands can be well classified as a
spectrum with $SU(3)$ symmetry if the projected neutron
and proton BCS vacuum states are considered to be two independent
$SU(3)$ representations coupled through the
$Q_\nu$-$Q_\pi$ interaction.

This may be demonstrated by considering a model Hamiltonian  with
$SU(3)^{\nu}\otimes SU(3)^{\pi}\supset SU(3)^{\nu+\pi}$
dynamical symmetry expressed in the form  (see Eq.\
(3.107) in Ref.\ \cite{FDSM}):
\begin{equation}
\hat{H}=\chi_\nu^{\mbox{\scriptsize
eff}}\hat{C}_{su3}^{\nu} +\chi_\pi^{\mbox{\scriptsize
eff}}\hat{C}_{su3}^{\pi}
-\chi_{\nu\pi}^{\mbox{\scriptsize
eff}}\hat{C}_{su3}^{\nu+\pi} + A \hat{I}^2,
\label{su3h}
\end{equation}
where $\hat{C}_{su3}^{\tau}$ are the
$SU(3)^{\tau}$ ($\tau=\nu,\pi$) Casimir operators for
neutrons, protons, and the n--p coupled symmetry
$SU(3)^{\nu+\pi}$ ($\tau=\nu+\pi$). The eigenvalue of
the lowest-order $SU(3)$ Casimir operator for a given
representation $(\lambda, \mu)$ is $C(\lambda, \mu)
={\small\frac12}
(\lambda^2+\mu^2+\lambda\mu+3\lambda+3\mu)$. We assume
that the two BCS vacua correspond to the $SU(3)$
symmetric representations $(\lambda_{\nu},0)$ and
$(\lambda_{\pi},0)$, respectively, and that the
permissible irreps $(\lambda, \mu)$ of $SU(3)^{\nu+\pi}$
are given by the Littlewood rule, $(\lambda, \mu) =
(\lambda_m-2\mu,\mu)$, where $\lambda_m = \lambda_\nu +
\lambda_\pi$ is the maximum value of $\lambda$ and
$\mu=0,1,2,\dots,\lambda_\pi$, if $\lambda_\pi \leq
\lambda_\nu$.

The eigenvalue spectrum $E[(\lambda, \mu) I]$ of
the Hamiltonian of (\ref{su3h}) obtained from the
states with angular momentum $I$ belonging to the
coupled representation
$\{ [\lambda_\nu, 0]\otimes[\lambda_\pi, 0] \} (\lambda,
\mu) I$ is
\begin{eqnarray}
E[(\lambda, \mu)
I]-E_{\mbox{\scriptsize g.s.}} &=&
\chi_{\nu\pi}^{\mbox{\scriptsize eff}} \left
[C(\lambda_m,0)-C(\lambda = \lambda_m-2\mu,\mu) \right ] + A I(I+1)
\nonumber \\
&=&\mu\hbar\omega_{\infty}\left
[1-\frac{\mu-1}{\lambda_m} \right ] + A I(I+1),
\label{su3e}
\end{eqnarray}
with
\begin{equation}
\hbar\omega_{\infty}~=~\frac{3}{2}
\lambda_m \chi_{\nu\pi}^{\mbox{\scriptsize
eff}}.
\end{equation}
The second term in Eq. (\ref{su3e}) gives a
rotational band with a moment of inertia parameter $A
= \hbar^2/2{\cal \Im}$. $E_{\mbox{\scriptsize g.s.}}$ is
the ground state energy corresponding  to the coupled
$SU(3)$ representation $[\lambda_m, 0]$
\begin{equation}
E_{\mbox{\scriptsize
g.s.}}=\chi_\nu^{\mbox{\scriptsize
eff}}C(\lambda_{\nu},0) +\chi_\pi^{\mbox{\scriptsize
eff}}C(\lambda_{\pi},0)
-\chi_{\nu\pi}^{\mbox{\scriptsize eff}}C(\lambda_m,0).
\label{su3eigen}
\end{equation}
The parameter $\lambda_m$ fixes the
$SU(3)$ representation $[\lambda_m, 0]$ of the g-band.
The parameter $\mu$ then fixes the representation
$[\lambda = \lambda_m-2\mu, \mu]$ of the excited bands
through the Littlewood rule.

A more physical meaning of the parameter $\mu$ emerges
from the first term of Eq. (\ref{su3e}), which  represents a
vibration-like spectrum which tends to be harmonic for
$\lambda_m \gg \mu$. In this limit, $\mu$ can be
interpreted as the number of phonons of this
``vibration" and $\hbar\omega_\infty$ as the one-phonon
excitation energy as $\lambda_m \rightarrow \infty$,
with its value being, approximately, the energy of the
first excited $I = 1^+$ state relative to the ground
state. The allowed angular momenta $I$ belonging to a
representation $[\lambda,\mu]$ are determined from the
usual $SU(3)$ subgroup reduction rules \cite{FDSM}. For
example, the representation $(\lambda, \mu) = (40,4)$
can have a $K=0$ band with $I=0,2,4, \ldots, 44$, a
$K=2$ band with $I=2,3,4, \ldots, 43$ and a $K=4$ band
with $I=4,5,6, \ldots, 41$ \cite{Vergados}.

The PSM calculated spectrum shown in Fig. 1 can then be
described by the $SU(3)$ symmetry spectrum of Eq.
(\ref{su3e}). The parameters $\lambda_m$,
$\hbar\omega_\infty$ and $A$ of Eq. (\ref{su3e}), which
fit best the PSM spectrum of Fig. 1, are
$\lambda_m=48$, $\hbar\omega_{\infty}=2.9$ MeV
($\chi_{\nu\pi}^{\mbox{\scriptsize eff}}=0.0403$), and
$A=0.013$ MeV. The $SU(3)$ band structure obtained with
these parameters is shown in Fig.\ 1 as the dashed lines
and reproduces remarkably well the PSM spectrum obtained from
numerical projection and diagonalization. The fact
that the parameter $\lambda_m=48$ fits the spectrum in
Fig. 1 suggests that the g-band of PSM may be associated
with the $SU(3)$ representation $[\lambda_m,0] =
[48,0]$. All the states shown in Fig. 1 can be labeled by
the $SU(3)$ irrep labels
$(\lambda,\mu)$ and the bandhead of each rotational band
within the irrep is labeled by the $SU(3)$ quantum
number $K$. Degeneracy at each spin may be deduced by
counting one for each $K$ band, except only even spins
are present for $K=0$ bands. We list the degeneracy of
the $(40,4)$ representation as an example in Fig.\ 1.
Not all PSM states can be seen clearly in the plot
because of the high degeneracy, but there is a
one-to-one correspondence between {\em all} predicted $SU(3)$
states and those observed in the PSM calculation.

One can see from Eq.\ (\ref{su3e}) and the $SU(3)$
reduction rules that the whole spectrum of Fig.\ 1 can
be viewed as a set of rotational bands built on
different multi-phonon excitation states with a phonon
energy $\hbar\omega=\hbar\omega_{\infty}\left
[1-\frac{\mu-1}{\lambda_m} \right ]$ and phonon spin $1\hbar$.
For example, a 3-phonon system could have two allowed
states with energy
$3\hbar\omega$ and total spin 1$\hbar$ and 3$\hbar$; a
4-phonon system could have three states with energy
$4\hbar\omega$ and total spin 0$\hbar$, 2$\hbar$, and
4$\hbar$; and so on. This provides an alternative
explanation of the degeneracy of the bands obtained by
the PSM diagonalization. Comparing the $SU(3)$ and
phonon classifications, one can see clearly that the
$SU(3)$ quantum numbers $\mu$ and $K$ in Fig.\ 1 may be
interpreted as the number of phonons and their allowed
total spins, respectively.

As long as $\hbar\omega_{\infty}$ is held constant, the
spectrum is sensitive to the effective $SU(3)$ quantum
number $\lambda_m$ only through a small anharmonicity. For
the present example, the phonon energy decreases
smoothly from 2.9 MeV to 2.5 MeV as the phonon number
increases from 1 to 7. If $\lambda_m\rightarrow\infty$, then
$\hbar\omega\rightarrow\hbar\omega_{\infty}$ and the
vibration becomes harmonic. Thus, the anharmonicity originates
in the finite number of particles for the
nuclear system.

\section{Further Inspection on the Wave Functions}

The remarkably quantitative agreement of all calculated excited states with the
$SU(3)$ spectrum demonstrated above makes it essentially certain
that an $SU(3)$ symmetry emerges from the purely numerical PSM calculation.
To remove all doubt, we may study the wave
functions. In this section, we shall do this in three
steps: we first compare angular momentum components in
the PSM and $SU(3)$ intrinsic states. Then, we
directly calculate overlaps of the actual wave functions
from the PSM calculation with $SU(3)$ wave functions.
Finally, we compute $B(E2)$ values from PSM calculations
that connect states in each band and compare with those
of the $SU(3)$ model.

\subsection{SU(3) Structure of the Intrinsic States}

Let us start by recalling that an intrinsic state
$|0_\tau \rangle$, $(\tau = \nu, \pi$), is related to a
projected state $|I_\tau \rangle$ by
\begin{equation}
|0_\tau \rangle ~=~ \sum_{I_\tau} C_{I_\tau} |I_\tau
\rangle.
\label{E1}
\end{equation}
In Eq.\ (\ref{E1}),
$C_I$ is the norm matrix element in the PSM
described in Eq.\ (2.33) of Ref.\ \cite{HS.95}
\begin{equation}
|C_I|^2 = \langle 0| \hat P^I |0
\rangle.
\end{equation}
It gives the probability of finding
angular momentum $I$ in the intrinsic state $|0 \rangle$ (with
$\sum_I |C_I|^2 = 1$, a sum rule obtained in Eq.\ (A80)
of Ref.\ \cite{HS.95}). In other words, the quantity
$|C_I|^2$ describes the angular momentum distribution in
the intrinsic state $|0 \rangle$. For the present
calculation, we projected the states with angular
momenta $I_\nu=0,2,...,32$ and $I_\pi=0,2,...,16$ from
the Nilsson-BCS intrinsic states $|0_\tau \rangle$ of
$^{168}$Er. The probabilities $|C_{I_\tau}|^2$ for
$I_\nu > 32$ and $I_\pi > 16$ were very small and hence
such states were not included in the basis state
$|[I_\nu \otimes I_\pi]I \rangle$ used in the present
PSM diagonalization. The values $|C_{I_\tau}|^2$ (for
the nucleus $^{168}$Er) calculated for the first nine
values $I_\tau=0,2,...,16$ contained in the intrinsic
state $|0_\tau \rangle$ are listed in Table 1. For
comparison, the values $|C_I|^2$ for the state $|I\rangle$
in the total intrinsic vacuum state $|0_\nu\rangle|0_\pi\rangle$
are also presented
in the same table.

In order to make a connection with the emergent $SU(3)$
symmetry at the intrinsic state level, we compare
the probabilities $C_{I_\tau}^2$(PSM) of the PSM
intrinsic states $|0_\tau \rangle_{\mbox{\scriptsize
PSM}}$ with the corresponding probabilities
$C_{I_\tau}^2$($SU(3)$) in the intrinsic states $|0_\tau
\rangle_{SU(3)}$ belonging to axially symmetric $SU(3)$
representation $[\lambda_\tau, 0]$. The latter
probabilities are given by Elliott's $a((\lambda, 0) I)$
coefficients \cite{Elliott}. Analytical formulas for the
$a((\lambda, \mu) I)$ are given in Table 2A of the
Vergados's paper \cite{Vergados}. The effective $SU(3)$
representation $[\lambda_\tau, 0]$ for protons and
neutrons is determined by equating the probabilities
$C_{I_\tau}^2$(PSM) = $C_{I_\tau}^2$($SU(3)$) =
$a((\lambda_\tau, 0) I_\tau)$ for a given $I_\tau$ (say
$I_\tau = 0$). For our present case of $^{168}$Er, the
values of $\lambda_\tau$ obtained by this procedure are
$\lambda_\nu \approx$ 32 and $\lambda_\pi \approx$ 16,
to the nearest even integer.

The coefficients $a((\lambda_\tau, 0) I)$ are then
calculated for all values of $I_\tau = 0, 2, ...,
\lambda_\tau$, allowed by the effective $\lambda_\tau$.
(We denote these values as $a_I$ in the following
discussion.) The $a_{I_\tau}$ values are compared with
the corresponding $C_{I_\tau}$ values of the PSM in Fig.\ 2.
The excellent agreement between these values shows that
the distribution of angular momenta of neutrons and
protons in our Nilsson--BCS vacuum states are very
similar to those in the intrinsic states of
the $SU(3)$-representations $[\lambda_\nu=32, 0]$ and
$[\lambda_\pi=16, 0]$.

The total Nilsson--BCS vacuum state $|0
\rangle_{\mbox{\scriptsize PSM}} = |0_\pi
\rangle_{\mbox{\scriptsize PSM}}|0_\nu
\rangle_{\mbox{\scriptsize PSM}}$ should then have an
effective
$SU(3)$-representation
$[\lambda=\lambda_\pi+\lambda_\nu, 0] = [48, 0]$ and the
probabilities $C_I^2$ of finding angular momentum $I$ in the
total PSM intrinsic state should be similar to the
probabilities $a((\lambda=48, 0) I)$ corresponding to
the $SU(3)$ intrinsic state of representation [48, 0]. The
agreement between these two probabilities is also shown
in Fig.\ 2.

These PSM effective $SU(3)$ representations for the
Nilsson--BCS intrinsic states of neutrons and protons for
$^{168}$Er in the three-shell valence space should be
compared with the effective representations
$[\lambda^{\mbox{\scriptsize eff}}_\nu, 
\mu^{\mbox{\scriptsize eff}}_\nu] = [40,0],
[\lambda^{\mbox{\scriptsize eff}}_\pi, 
\mu^{\mbox{\scriptsize eff}}_\pi] = [24,0]$, and
$[\lambda^{\mbox{\scriptsize eff}}, 
\mu^{\mbox{\scriptsize eff}}] = [64,0]$ obtained
by Kahane {\it et al.} \cite{kahane} for just the
Nilsson intrinsic state of $^{168}$Er in the $\nu
(82-126)$ and $\pi(50-82)$ single major shell valence
space. These representations are larger than the
effective representations $[\lambda_\nu=32, 0]$ and
$[\lambda_\pi=16, 0]$ mainly because Kahane {\it et al.}
have included all the projected states with probabilities
$|C_I|^2 < 0.001$, which have been omitted in the present
PSM calculations.
For comparison, we also note that in the
one major shell valence space, the pseudo-$SU(3)$
representations for the normal parity nucleons in
$^{168}_{68}$Er$_{100}$ are \cite{pseudo}
$[\lambda^{\mbox{\scriptsize pseudo}}_{\nu},
\mu^{\mbox{\scriptsize pseudo}}_\nu] = [20, 4]$, 
$[\lambda^{\mbox{\scriptsize pseudo}}_{\pi},
\mu^{\mbox{\scriptsize pseudo}}_\pi] = [10, 4]$, and
$[\lambda^{\mbox{\scriptsize pseudo}}_{n}=30,
\mu^{\mbox{\scriptsize pseudo}}_{n}=8]$.
These representations are strongly triaxial
but their effective axially symmetric representations
would be $[\lambda^{\mbox{\scriptsize eff}}_\nu=24,0]$,
$[\lambda^{\mbox{\scriptsize eff}}_\pi=14,0]$ and
$[\lambda^{\mbox{\scriptsize eff}}=38,0]$.
In the FDSM
\cite{FDSM}, the $SU(3)$ representation for 10 neutrons
in the normal parity states of the (82 -- 126) shell is
$[\lambda^{\mbox{\scriptsize FDSM}}_{\nu},
\mu^{\mbox{\scriptsize FDSM}}_\nu] = [10, 0]$. 
There are also 10 protons in normal parity states 
of the (50 -- 82) shell, and they do not show an $SU(3)$,
but an $SO(6)$ symmetry that corresponds to a $\gamma$-soft rotor.
However, when the protons are
strongly coupled to the neutrons, they will be
synchronized with neutrons and stabilized at $\gamma\approx 0$, 
so that the proton system behaves like
an $SU(3)$ rotor.  Therefore, an effective $SU(3)$ representation
$[\lambda_{\pi}^{\mbox{\scriptsize FDSM}}, 
\mu_{\pi}^{\mbox{\scriptsize FDSM}}] = [10, 0]$ could be assigned to the
protons.  Quantitatively, the FDSM with one major shell can only fit to
the first three bands in Fig. 1. 
Larger deviations are expected for the higher bands, 
although the qualitative band structure remains correct.
If the $SU(3)$
symmetry which appears to emerge from the PSM
calculation presented here is identified as the FDSM
symmetry in the extended three-shell valence space 
with the effective
neutron and proton numbers $n_\nu^{\mbox{\scriptsize eff}}=32$ 
and $n_\pi^{\mbox{\scriptsize eff}}=16$, then 
agreement with the PSM results is perfect, as shown in Fig. 1.

\subsection{SU(3) Symmetry in the Wave Functions}

Next we shall verify that the structure of the PSM
wave function $|I \rangle_{\mbox{\scriptsize PSM}}$
belonging to different bands obtained by diagonalization
is similar to that of the $SU(3)$ wave function
$|(\lambda,\mu),I \rangle_{SU(3)}$ belonging to the
$SU(3)$ representation $[\lambda,\mu]$ obtained by
coupling the $[\lambda_\pi,0]$ and $[\lambda_\nu,0]$.
Here we consider only the g-band. The PSM wave functions
for the g-band obtained by diagonalizing the Hamiltonian
in the basis $\{(I_\nu \otimes I_\pi) I\}$ can be
written as
\begin{equation} |I
\rangle_{\mbox{\scriptsize PSM}} ~=~
\sum_{I_\nu,I_\pi} f_{\mbox{\scriptsize PSM}} (I_\nu
I_\pi ; I) |(I_\nu \otimes I_\pi) I \rangle,
\label{PSMwave}
\end{equation}
and the total
$SU(3)$ wave function for the g-band representation
$[\lambda,0]$ can be written as
\begin{eqnarray}
|(\lambda,\mu),I \rangle_{SU(3)} &=&
\aleph_I \sum_{I_\nu,I_\pi} a_{I_\nu} a_{I_\pi} \langle
I_\nu 0 I_\pi 0 | I 0 \rangle |[I_\nu \otimes I_\pi] I
\rangle
\nonumber\\
&=& \sum_{I_\nu,I_\pi} f_{SU(3)} (I_\nu
I_\pi ; I) |(I_\nu \otimes I_\pi) I \rangle,
\end{eqnarray}
where
$\aleph_I$ is the normalization constant
\begin{equation}
|\aleph_I|^2 = { 1\over {\sum_{I_\nu,I_\pi} ( a_{I_\nu}
a_{I_\pi} \langle I_\nu 0 I_\pi 0 | I 0 \rangle)^2}} .
\end{equation}
The overlap of these two wave functions
is
\begin{eqnarray}
O_I &=& _{SU(3)}
\langle (\lambda,\mu),I | I \rangle_{\mbox{\scriptsize
PSM}} \nonumber\\ &=& \sum_{I_\nu,I_\pi} f_{SU(3)}
(I_\nu I_\pi ; I) f_{\mbox{\scriptsize PSM}} (I_\nu
I_\pi ; I).
\end{eqnarray}

Knowing the $SU(3)$ coefficients $f_{SU(3)}$ from
$a_{I_\nu}$ and $a_{I_\pi}$, and the PSM coefficients
$f_{\mbox{\scriptsize PSM}}$ from numerical
diagonalization, we are able to compute their overlap.
The calculated overlaps $O_I$ are listed in Table 2
\footnote{A similar calculation for the overlaps for the
PSM wave functions $|I \rangle_{PSM}$ belonging to
higher bands with the $SU(3)$ wave functions of the
representations [44,2], [40,4], {\it etc.} requires
calculation of $SU(3)$ Clebsch--Gordon coefficients. The proof should
be straightforward and is not given here.}. All the
numbers are very close to 1, indicating a strong overlap
of the PSM wave functions with the $SU(3)$ wave
functions.

\subsection{SU(3)-like Systematics of $B(E2,
I\rightarrow I-2)$ values}

The $B(E2)$ values are useful to determine the
collectivity of a band. In the discussion of the last
section, the numerically calculated PSM states were
classified using group theoretical methods.
We shall now verify this classification with $B(E2)$
calculations using PSM wave function, which will allow us to classify
PSM states into collective bands.
Since we have proved the similarity of
the PSM wave functions to the $SU(3)$ ones, we expect
to obtain similar $B(E2)$ values if we make a corresponding
calculation using $SU(3)$ wave functions.

The $B(E2)$ transitions are calculated as
\begin{equation}
B(E2, I_i\rightarrow I_f) ~=~
{{|\langle I_f \parallel {\cal T}^2 \parallel I_i
\rangle |^2}\over {2I_i+1}},
\label{BE2}
\end{equation}
where the operator ${\cal T}^2 = {\cal
T}^2_\nu + {\cal T}^2_\pi$ is related to the quadrupole
operators by
\begin{eqnarray}
{\cal T}^2_\nu &=&
e^{\mbox{\scriptsize eff}}_\nu \sqrt{5\over {16\pi}}
Q^2_\nu,
\nonumber\\
{\cal T}^2_\pi &=&
e^{\mbox{\scriptsize eff}}_\pi \sqrt{5\over {16\pi}}
Q^2_\pi.
\end{eqnarray}
In the calculations, we have
used the usual effective charges $e^{\mbox{\scriptsize
eff}}_\nu = 0.5e$ and $e^{\mbox{\scriptsize eff}}_\pi =
1.5e$. By employing the PSM wave functions of Eq.
(\ref{PSMwave}), the reduced matrix element appearing in
Eq. (\ref{BE2}) can be evaluated as
\begin{eqnarray}
\langle I_f \parallel {\cal T}^2 \parallel I_i \rangle
&=& \sum_{I_f(I_{f,\nu}, I_{f,\pi})}
\sum_{I_i(I_{i,\nu}, I_{i,\pi})} f_f(I_{f,\nu}
I_{f,\pi}; I_f) f_i(I_{i,\nu} I_{i,\pi}; I_i)
\nonumber\\
& & \langle (I_{f,\nu} \otimes I_{f,\pi})
I_f \parallel {\cal T}^2_\nu + {\cal T}^2_\pi \parallel
(I_{i,\nu} \otimes I_{i,\pi}) I_i \rangle ,
\end{eqnarray}
in which the matrix elements for a
coupled system can be explicitly expressed as
\cite{BMbook}
\begin{eqnarray}
\langle [I_{f,\nu} \otimes I_{f,\pi}] I_f & &\parallel
{\cal T}^2_\nu \parallel [I_{i,\nu} \otimes I_{i,\pi}]
I_i \rangle ~=~ \sqrt{(2I_i+1)(2I_f+1)}
\nonumber\\ & &
(-)^{I_{f,\nu}+I_{f,\pi}+I_i} {\cal W}(I_{i,\nu}
I_{i,\pi} 2 I_f; I_i I_{f,\nu}) \langle I_{f,\nu}
\parallel {\cal T}^2_\nu \parallel I_{i,\nu} \rangle
\delta_{I_{f,\pi} I_{i,\pi}},
\nonumber\\
\langle
[I_{f,\nu} \otimes I_{f,\pi}] I_f & &\parallel {\cal
T}^2_\pi \parallel [I_{i,\nu} \otimes I_{i,\pi}] I_i
\rangle ~=~ \sqrt{(2I_i+1)(2I_f+1)}
\nonumber\\ & &
(-)^{I_{i,\nu}+I_{i,\pi}+I_f} {\cal W}(I_{i,\nu}
I_{i,\pi} I_f 2; I_i I_{f,\pi}) \langle I_{f,\pi}
\parallel {\cal T}^2_\pi \parallel I_{i,\pi} \rangle
\delta_{I_{f,\nu} I_{i,\nu}}.
\end{eqnarray}

In the $SU(3)$ model, assuming the transition operator
${\cal T}^2=\alpha P^2$, where $P^2$ is the generator of
the coupled $SU(3)^{\nu+\pi}$, the $B(E2)$ formula for the
intra-band transition is
\begin{equation}
B(E2,
I_i\rightarrow I_f) ~=~ \frac{2I_f+1}{2I_i+1} \alpha^2
\left |\langle (\lambda\mu)KI_i,(11)2 \parallel
(\lambda\mu)KI_f \rangle \right |^2 C(\lambda,\mu),
\label{SU3BE2}
\end{equation}
where $\langle (\lambda\mu)KI_i,(11)2
\parallel (\lambda\mu)KI_f \rangle$ is the
$SU(3)\supset R(3)$ Wigner coefficient, and $\alpha$ is
a parameter of the effective transition operator in the
$SU(3)$ model determined by fitting data. The inter-band
transition rate is zero.  In the symmetry limit,
if the transition operator is not a generator of
$SU(3)^{\nu+\pi}$ or there exists small symmetry
breaking in the wavefunction, the inter-band transition rate would be
expected to be small but not zero.

Generally, the shell-model $E2$ operator has different
effective charges for
neutrons and protons and as such is not a generator of the
coupled $SU(3)^{\nu+\pi}$. We note,  
however, that Eq. (\ref{SU3BE2}) is still a good approximation for
this more general $E2$ operator through the inclusion of a
factor $(e_\nu\lambda_\nu+e_\pi\lambda_\pi)/(\lambda_\nu+\lambda_\pi)$
in the proportionality constant $\alpha$. Different neutron and proton
effective charges (or gyromagnetic ratios for $M1$'s) will also
give rise to inter-band transitions, which is an essential mechanism
to obtain the strong $M1$'s from scissors states to ground band.

The $B(E2)$ values calculated by the PSM are shown in Fig.\
3, in comparison with the $SU(3)$ results and the
experimental data \cite{BE2}. For the g-band (labeled  as the
first $0^+$ band in the tables), good agreement is found
for all states except for the transition $6^+
\rightarrow 4^+$. Besides the g-band, we have
calculated the $B(E2)$'s for the first $1^+$ band, the second
$0^+$ band, and the first $2^+$ band
(see Fig.\ 1). The values displayed in Table 3
confirm strong collectivity in each of these
intra-transition bands and is in good agreement with
what found in the $SU(3)$ model.
Linking $B(E2)$ values between any of two bands are also
calculated, with the results displayed  in Table 4. These
inter-band $B(E2)$ values are typically two orders of
magnitude smaller than the values for the intra-band
transitions. From
these calculations, the sets of nearly degenerate bands seen
in the PSM calculations of Fig.\ 1 (for example, the
second $0^+$ band and the first $2^+$ band) can be
separated easily. All these results strongly support
the conclusion that the structure
of well deformed nuclei resulting from PSM calculations
indeed possesses
a very strong $SU(3)$ symmetry.

\section{Comments on the Emergent SU(3) Symmetry}

The connection between $SU(3)$ symmetry and nuclear rotation has
a long history. It was first explored by Elliott in his classic
1958 papers \cite{Elliott} for $sd$-shell nuclei. This idea was
later extended to heavy systems using  the pseudo-$SU(3)$ model
\cite{pseudo}, which showed that it is possible to represent
approximately the ground band of a deformed nucleus by one leading
$SU(3)$ representation belonging only to the normal parity
nucleons in a single major shell. However this model has
difficulty in reproducing $\beta$-bands of heavy nuclei, and
quantitative results generally do not appear in simple symmetry
limits and therefore require numerical calculations. Later, the
interacting boson model (IBM) \cite{IBM1} demonstrated that
nuclear rotational motion including $\beta$- and $\gamma$-bands
can be described by an $SU(3)$ dynamical symmetry based on
$s$-$d$ or $s$-$d$-$g$ bosons. The fermion dynamical symmetry
model (FDSM), based on the $S$-$D$ fermion pairs of normal parity
nucleons in a major shell, demonstrated analytically the
equivalence between the particle--rotor model and the $SU(3)$
dynamical symmetry limit of the FDSM when particle number
$n\rightarrow\infty$; if the Pauli effect is neglected (by
assuming the shell pair-degeneracy $\Omega\rightarrow\infty$) the
FDSM reduces to the IBM (see Section 3 and 4 in Ref.\
\cite{FDSM}).

However, all the above-mentioned
models are algebraic and the possibility for
symmetries
arises naturally in them. The
extended PSM described here has not built an explicit
$SU(3)$
symmetry into the problem, and no free parameters
have been adjusted. Nevertheless, the spectra and wavefunctions obtained from
the
PSM calculations
can again be well classified
using the representation theory of the
$SU(3)$ group. The $SU(3)$ symmetry just
{\em emerges naturally} at the macroscopic level
from a shell model diagonalization in the
basis obtained by angular momentum projection from a
deformed intrinsic state. This is a non-trivial result
because our study is not confined in the g-band, but
extended up to high spins and high excitations, and because the
correspondence has been demonstrated not only at the spectral level, but in the
wavefunctions and electromagnetic transition rates too.  This
strongly indicates that a well deformed nucleus is in
fact a very good $SU(3)$ rotor.

Now we may ask an important question: what is the nature of the
$SU(3)$ symmetry that emerges from the PSM? There are many
$SU(3)$ models. Mathematically, they are equivalent {\em if
their representations are the equivalent}. Physically, the models
differ in the microscopic basis on which the symmetry is
built, and have different permissible $SU(3)$ irreps, and thus
different band structures. The emergent $SU(3)$ symmetry we have found
here is in principle consistent with any $SU(3)$ symmetry model
that has allowed representations consistent with the observed states and band
structure.  We emphasize that this is a {\em physical} rather than mathematical
criterion, for it is the physical input to an $SU(3)$ model that determines its
allowed irreps.
Therefore, to answer the
question of which $SU(3)$ model is consistent with the
symmetry of the  well deformed rotor that we have explored in
this paper,  one  should look at the entire band structure
and not just the ground-state  rotational band.

As we have already demonstrated, the $SU(3)$ symmetry
that emerges from the PSM for $^{168}$Er corresponds to
effective $\{
[\lambda_\nu=32, 0]\otimes[\lambda_\pi=16, 0]
\} (\lambda, \mu) I$ irreps for neutrons and protons, respectively.
Among the $SU(3)$ models being employed in nuclear structure physics,
the FDSM (or equivalently the IBM that results if Pauli effects are
neglected in the FDSM) can naturally accommodate such a representation
if the shell degeneracy
$\Omega$ is large enough \cite{FDSM}.  As can be easily
checked, no other currently existing $SU(3)$ fermion model permits naturally
such an $SU(3)$ representation.  For
example,  the leading $SU(3)$ irreps for $^{168}$Er given by the
pseudo-$SU(3)$ model \cite{pseudo} would be
$\{ [\lambda^{\mbox{\scriptsize pseudo}}_{\nu}=20,
\mu^{\mbox{\scriptsize pseudo}}_\nu=4]\otimes
[\lambda^{\mbox{\scriptsize pseudo}}_{\pi}=10,
\mu^{\mbox{\scriptsize pseudo}}_\pi=4]
\} (\lambda, \mu) I$, which will give a band
structure very different from what we have seen in
Fig.\ 1.  Thus, we conclude that in a model such as the pseudo-$SU(3)$,
the $SU(3)$ spectrum found in the numerical PSM calculation cannot be
reproduced by any simple symmetry limit and could emerge (if at all) only
from a large mixing of irreps through numerical diagonalization.

However, it should be noted that although we have demonstrated
that the $SU(3)$ symmetry that emerges from the PSM is of
FDSM-type (or the IBM-type if Pauli effects are omitted),
this does not necessarily
mean that the simplest single-major shell FDSM can accommodate such an
$SU(3)$ symmetry microscopically. If we assume that
the $SU(3)$ symmetry which emerges for
$^{168}$Er in the present PSM calculation is the fermion
dynamical $SU(3)$ symmetry of Ref.\ \cite{FDSM} based on $S$-$D$
fermion pairs, the $SU(3)$ quantum numbers $\lambda_\nu$ and
$\lambda_\pi$ have a very simple interpretation.  They are the
effective number of neutrons and protons in the valence space,
$\lambda_\nu=n^{\mbox{\scriptsize eff}}_{\nu}$ and
$\lambda_\pi=n^{\mbox{\scriptsize eff}}_{\pi}$, which form the
coherent $S$-$D$ pairs and are responsible for the collective
rotation. In the simplest implementations of the
FDSM, the model space is restricted to
one major shell for protons and neutrons. The effective neutron
(proton) number $n^{\mbox{\scriptsize eff}}_{\nu}$
($n^{\mbox{\scriptsize eff}}_{\pi}$) is just the number of
neutrons (protons) in the normal-parity levels of a single major
shell, which is obviously too small to satisfy the requirement of
$n^{\mbox{\scriptsize eff}}_{\nu}\geq 32$ and
$n^{\mbox{\scriptsize eff}}_{\pi}\geq 16$ for the $^{168}$Er
case.  In order to accommodate the $\{ [\lambda_\nu=32,
0]\otimes[\lambda_\pi=16, 0] \} (\lambda, \mu) I$ $SU(3)$
symmetry, the one-major shell FDSM has to be extended to a
multi-major shell FDSM. Namely, the coherent $S$-$D$ pairs should
be redefined in a multi-major shell space. This has already been
discussed extensively in conjunction with the extension of the
FDSM to the description of superdeformation, and much of that
discussion applies to the present case.

One should note in this regard that there is a conceptual
distinction between the $SU(3)$ symmetry that emerged from the
PSM diagonalized in multiple shells and that which arises in the
one-major shell FDSM.  In the PSM calculations the $SU(3)$
symmetry arises from the explicit dynamical participation of both
normal and abnormal-parity nucleons, while in the symmetry limit
of the single one-major shell FDSM the $SU(3)$ symmetry itself
arises entirely from the normal-parity nucleons and the abnormal
parity orbital enters only implicitly through the Pauli effect
and by renormalizing the $SU(3)$ parameters. Thus, the
multi-major shell FDSM is also necessary to resolve  this
conceptual ambiguity. The reason that the one-major shell FDSM
treats the contribution of the abnormal-parity component to the
collective motion differently from normal parity orbitals is
because in a single major shell there is only one abnormal
parity  level.  It has been shown that a single-$j$ shell does
not have enough quadrupole collectivity (the possibility to form a
$D$ pair in the $i_{13\over 2}$ intruder level compared to that
in the corresponding normal-parity levels is only $6\over 46$;
see Ref.\ \cite{def}). When the FDSM is extended to a three-shell
valence space as in the PSM, the situation will be changed.  A
bunch of abnormal-parity levels, which are located just below the
normal-parity levels, will open up. Abnormal-parity nucleons thus
have enough collectivity to form coherent $D$ pairs and
participate in collective motion.  This idea has already been
developed in the extension of the FDSM to superdeformation.

Although the basic theme of the present work is that
the $SU(3)$ symmetry-like features emerge for deformed nuclei
from a realistic Hamiltonian which has no such symmetry
at a microscopic level, it is tempting to look for
the microscopic basis underlying the emergent $SU(3)$ symmetry.
The preceding arguments give us strong reason to  speculate that
this emergent $SU(3)$ symmetry is just a manifestation of the FDSM
operating over a three-shell valence space.

\section{Comments on the Scissors Mode Vibrations}

A second important consequence of our study is that
the rotational states that we have just described in terms of an $SU(3)$
symmetry
can be regrouped and interpreted as phonon
vibrations (even though
no explicit vibrational information has been given to the
calculation).
A standard signature of ideal harmonic
vibrational motion is the appearance  of an equally spaced
spectrum of energy levels with a characteristic degeneracy pattern.
A similar connection between
$SU(3)$ symmetry and vibration in rotating nuclei
was pointed out by Wu {\em et al.} for $\beta$-
and $\gamma$-vibrations \cite{wusu3}. In the present case, we
have clearly shown that the band heads of the different
$SU(3)$ representations $[\lambda, \mu]$ obtained by
coupling the proton--neutron representations
$[\lambda_\pi, 0]$ and $[\lambda_\mu, 0]$ tend to be
equally spaced. We have also shown that the spectra of
the states obtained by PSM diagonalization
is dominated by nearly equally spaced bands (apart from a small
anharmonicity). Therefore it is appropriate to call
these ``vibrational bands". Physically the only
vibrational mode allowed for protons and neutrons that
retain their individual $SU(3)$ irreps
$[\lambda_\tau, 0]$
(this algebraic restriction is related to
a geometrical requirement of fixed deformation for both neutrons and protons)
is the ``scissors mode".  The scissors mode corresponds geometrically
to an oscillation in the relative orientation angle of the quadrupole-deformed
neutron and proton potentials.

Traditionally, collective vibrations appear as a
consequence of small oscillations around the equilibrium
in the deformed potential, as we know from the physics
of $\beta-$ and $\gamma-$vibrations \cite{BMbook}. Theories of the
RPA type allow superposition of states displaced around
the equilibrium, thus describing the physics of small
oscillations, as required for $\beta-$ and
$\gamma-$vibrations. However, the collective
vibrations considered here originate from relative motion
between proton and neutron systems at fixed
deformation. As long as one does not decompose the
neutron and proton degrees of freedom, the physics
discussed in our work is not contained in such  theories.

On the other hand, it is clear that the present results do not
describe $\beta$- and $\gamma$-spectra because
the $\beta$- and $\gamma$-vibrational modes do not
belong to the present configuration space. This can be
seen clearly with the aid of the preceding $SU(3)$
classification of the PSM states. The $\beta$- and
$\gamma$-vibrations (the quadrupole phonon or spin-2$\hbar$ excitations),
correspond to a coherent superposition
of the $[(\lambda_{\nu}-2\mu,\mu)\otimes
(\lambda_{\pi},0)]$
and $[(\lambda_{\nu},0)\otimes
(\lambda_{\pi}-2\mu,\mu)]$
representations
with $\mu$ equal to a non-zero  even integer.
In other words,
either the neutron  or the proton core (or both), must be excited. It is
easy to show from Eq.\ (\ref{su3h}) that the dominant $SU(3)$
representation for the lowest $\beta$- and
$\gamma$-bands is $[(\lambda_{\nu}-4,2)\otimes
(\lambda_{\pi},0)] (\lambda-4,2)$
if $\lambda_{\nu} > \lambda_{\pi}$. The
$\beta$-band head energy is given by
\begin{equation}
E_{\beta 0}-E_{\mbox{\scriptsize
g.s.}} = -3\lambda_{\nu}\chi_\nu^{\mbox{\scriptsize eff}}
+2\hbar\omega_{\infty}, \hspace{24 pt}
\hbar\omega_{\infty}=\frac{3}{2}\lambda_m
\chi_{\nu\pi}^{\mbox{\scriptsize
eff}}
\end{equation}
Thus the $\beta$-vibrational band head energy is significantly
lower than the energy $E_{\mbox{\scriptsize 2nd}-0^+}
 = 2\hbar\omega_{\infty}$ of the
lowest $K=0^+$ band of the scissors mode excitation shown in
Fig.\ 1 which corresponds to the
$[(\lambda_\nu, 0)\otimes (\lambda_\pi, 0)] (\lambda_m-4, 2)$
representation. In fact, $E_{\beta 0}$ will be lower than
even the energy $E_{\mbox{\scriptsize
1st}-1^+}$ of the band head of the first 1$^+$ scissors mode
$E_{\mbox{\scriptsize
1st}-1^+} = \hbar\omega_{\infty} + 2A$,
if the condition $\chi_\nu^{\mbox{\scriptsize
eff}}>\frac{\lambda_m}{2\lambda_\nu}\chi_{\nu\pi} ^{\mbox{\scriptsize
eff}}$ is fulfilled.
In the case of $^{168}$Er discussed here, this condition becomes
$\chi_\nu^{\mbox{\scriptsize
eff}}>0.75\chi_{\nu\pi} ^{\mbox{\scriptsize
eff}}$.

This means that, unlike the scissors mode, the $\beta$- and
$\gamma$-vibrations do not correspond to  relative motion between
fixed neutron and proton fields,  but to an internal collective
excitation of neutrons or/and protons. This picture is consistent
with the conventional shape vibration picture \cite{BMbook}.
Therefore, to obtain the classical $\beta$- and $\gamma$-bands
within the present framework one must build a richer set of
correlations into the vacuum. In principle, this could be
achieved by mixing a large set of multi-quasiparticle states, but
in practice, the configuration space may be too large to handle
if there is substantial collectivity in these modes. A more
efficient way to introduce these states is to build those degrees
of freedom into the intrinsic basis. Very recently, $\gamma$-band
and multi-phonon $\gamma$-vibrational states in rare-earth nuclei
have been obtained by Sun {\it et al.} \cite{gamma} by using
three-dimensional angular momentum projection on triaxially
deformed potential \cite{TPSM}.

\section{The First 1$^+$, 2$^+$, and the Second 0$^+$
Bands}

Several lowest excited bands appearing in Fig.\ 1
warrant further discussion. In Fig.\ 4 we plot these
bands, together with the g-band from the PSM
calculations and from experiment \cite{BE2}.
There are other observed low-lying collective states
in this nucleus (for example, the 2$^+$ $\gamma$-band starting
at 0.821 MeV \cite{BE2}, which was extensively studied in the PSM framework
in Ref. \cite{gamma}). Discussion of these states
is beyond the scope of the present paper, and therefore,
we omit plotting them in Fig.\ 4.

The first 1$^+$ band corresponds to a 1-phonon
excitation with the excitation energy relative
to the ground state depending on the interaction strengths
used in the calculation. The good agreement of the
calculated g-band with data (and similar results for
many other calculations in this mass region
\cite{HS.95}) suggests that the strengths we use here
are realistic. These excitations are due to an $SU(3)$
coupling $(n_{\nu},0)\otimes (n_{\pi},0)$ in which both
neutron and proton intrinsic systems remain in the
ground states; thus they are related physically to
relative motion between neutrons and protons. We
conclude that this may be the 1$^+$ scissors mode band
previously suggested in other models
\cite{Rowe,Hilton,two.rotor,nojarov.86,IBM}.
Our obtained bandhead of this 1$^+$ band is at an excitation
of about 3 MeV, reasonably within the energy range of 
experimental observations for scissors mode 
\cite{Exp}.

The present results indicate that the PSM provides a microscopic
framework in which collective modes that may be closely
identified with those proposed in earlier geometrical and
algebraic descriptions emerge as the lowest excitations.
Furthermore, it is already well established that the PSM
describes structures built on quasiparticle excitation very well
\cite{HS.95}. Therefore, the extension of the present
calculations to a larger basis including 2 and possibly 4
quasiparticle excitations of the n--p coupled vacuum will provide
a microscopic formalism in which collective and quasiparticle
degrees of freedom enter on an equal footing. Such calculations
are possible and are presently being explored. We may expect that
the long-debated question of whether the observed $1^+$ states
are collective or two-quasiparticle in nature may then be
resolved through such quantitative calculations of this sort.

The other excited bands are generalizations of the scissors mode
corresponding to multi-phonon excitations. In Fig.\ 4,
the next two bands at 6.5 MeV excitation energy
are nearly degenerate 2-phonon states, corresponding to the coupling
of two 1$\hbar$-phonons to total spins
0$\hbar$ and 2$\hbar$. These are theoretically predicted
multi-phonon excitations of the scissors mode, and have not
to our knowledge
been seen experimentally. Although the level density is
expected to be high at that excitation energy and symmetry breaking
will fractionate the strength,
these predicted states might be detectable in the new
generation of modern detectors.

\section{Summary}

We have found many new collective states in a shell model
diagonalization based on separately projected neutron and proton
Nilsson + BCS vacuum states. We have shown that these states
exhibit an almost perfect $SU(3)$ symmetry, both in their spectra
and their wavefunctions. We have shown also that these states can
be classified systematically in a phonon spectrum with weak
anharmonicity. Among these states, the lowest 1$^+$ band at about
3 MeV corresponds to the scissors mode predicted in a classical
geometrical picture. The PSM is a shell model diagonalization
without explicit $SU(3)$ symmetries. However, the quantitative
agreement with an $SU(3)$ model provides an algebraic fermion
classification scheme for the states obtained from the PSM
diagonalization, and suggests that the projected BCS vacuum for a
well-deformed system is very close to the $SU(3)$ dynamical
symmetry limit of an $S$-$D$ pair fermion system. This in turn
implies a good boson algebraic symmetry if Pauli effects may be
ignored. Finally, we have proposed that the extension of the
present calculations to include quasiparticle excitations
provides a quantitative framework to determine whether ``scissors
mode'' 1$^+$ states are more properly viewed as collective
excitations or as quasiparticle states.

The present paper deals only with even-even nuclear system.
However, our preliminary results \cite{odd} have shown that
similar conclusions can also be drawn for odd- and odd-odd nuclei.

Y.S. thanks Professor Gui-Lu Long of 
Department of Physics of Tsinghua University
for warm hospitality, where the final version of this paper was completed.
K.H.B. thanks Dr. S. Raman of Oak Ridge National
Laboratory for partial support.

\baselineskip = 14pt
\bibliographystyle{unsrt}

\newpage

\begin{table}[h]
\begin{center}
\caption{The calculated angular momentum distribution in
the intrinsic vacuum states of Nilsson--BCS, $|C_I|^2$,
for the first nine angular momenta, for the neutron
vacuum $|0_\nu \rangle$, the proton vacuum $|0_\pi
\rangle$, and the product vacuum $|0 \rangle = |0_\nu
\rangle |0_\pi \rangle$. }
\begin{tabular}{|cccc|}
Spin I & $|C_{I_\nu}|^2$ & $|C_{I_\pi}|^2$ & $|C_{I}|^2$ \\
\hline 0&0.0313&0.0591&0.0203\\ 2&0.1419&0.2452&0.0954\\
4&0.2041&0.2877&0.1486\\ 6&0.2078&0.2158&0.1712\\
8&0.1695&0.1186&0.1647\\ 10&0.1161&0.0504&0.1381\\
12&0.0684&0.0171&0.1032\\ 14&0.0352&0.0047&0.0696\\
16&0.0159&0.0011&0.0426\\
\end{tabular}
\end{center}
\label{table1}
\end{table}

\begin{table}
\caption{The overlap of the PSM and the $SU(3)$ wave
functions. } \begin{tabular}{|cc|}
Spin $I$ & Overlap
$O_I$\\ \hline 0&0.9686\\ 2&0.9683\\ 4&0.9677\\
6&0.9671\\ 8&0.9668\\ 10&0.9669\\ 12&0.9673\\
14&0.9678\\ 16&0.9680\\ 18&0.9676\\
\end{tabular}
\label{table2}
\end{table}

\begin{table}
\caption{Calculated B(E2, $I \rightarrow I-2$) values
(in $e^2b^2$) for several excited bands with PSM wave
functions. } \begin{tabular}{|ccccc|}
Spin I & 1st $0^+$
& 1st $1^+$ & 2nd $0^+$ & 1st $2^+$ \\ \hline 2&1.104&
&1.009& \\ 4&1.578&1.249&1.383&0.599\\
6&1.739&1.541&1.469&1.196\\ 8&1.821&1.667&1.550&1.487\\
10&1.872&1.735&1.617&1.639\\ 12&1.906&1.778&1.661&1.718\\
\hline 3&   &0.917& &   \\ 5&   &1.459& &0.969\\ 7& &1.647&
&1.339\\ 9& &1.741& &1.512\\ 11&    &1.797& &1.609\\
\end{tabular}
\label{table3}
\end{table}

\begin{table}
\caption{Calculated B(E2, $I \rightarrow I-2$) values
(in $e^2b^2$) for linking transitions between bands. }
\begin{tabular}{|ccccc|}
Spin I & 1st $1^+ \to$ 1st
$0^+$ & 2nd $0^+ \to$ 1st $1^+$ & 1st $2^+ \to$ 2nd
$0^+$ & 1st $2^+ \to$ 1st $1^+$ \\
\hline 2&0.017& &0.007 &
\\ 4&0.020&0.016&0.063&0.017\\
6&0.020&0.026&0.094&0.009\\ 8&0.020&0.032&0.055&0.004\\
10&0.020&0.034&0.022&0.002\\ 12&0.019&0.035&0.008&0.001\\
\hline 3&   &   &   &0.021\\ 5& &   &   &0.024\\ 7&
&   &   &0.023\\
9&  &   &   &0.022\\ 11&    &   &   &0.021\\
\end{tabular}
\label{table4}
\end{table}

\newpage
\begin{figure} 
\caption{ Spectrum of collective excitations
corresponding to coupled rotation of neutrons and protons. Symbols
are calculated by the Projected Shell Model; dashed lines are calculated
by an $SU(3)$ Symmetry Model. For those symbols, ``$\bullet$" represents 
states having $K$ = 0 and 1, ``$\circ$" having $K$ = 2 and 3, 
``$-$" having $K$ = 4 and 5, and ``$\times$" having $K$ = 6 and 7. 
Many states are covered by other
states because of the high level of degeneracy. The degeneracy is
indicated explicitly for the (40,4) states. }
\label{figure.1}
\end{figure}

\begin{figure}
\caption{ Comparison of angular momentum distribution in the
intrinsic vacuum states of Nilsson--BCS calculated in the PSM and
the $SU(3)$ representation. The curves connect the PSM values and
the symbols represent the $SU(3)$ values. }
\label{figure.2}
\end{figure}

\begin{figure}
\caption{Comparison of the calculated B(E2, $I \rightarrow I-2$)
values from the PSM with experimental data for $^{168}$Er
g-band\protect\cite{BE2} and the $SU(3)$ limit (calculated from
Eq.\ (\protect\ref{SU3BE2}) with the parameter $\alpha=0.0683$).}
\label{figure.3}
\end{figure}

\begin{figure}
\caption{Spectrum of the ground band, the first 1$^+$ and 2$^+$,
and the second 0$^+$ collective bands in $^{168}$Er.}
\label{figure.4}
\end{figure}

\end{document}